\documentclass[aps,prl,twocolumn,reprint,superscriptaddress]{revtex4}
\usepackage{amsmath}
\usepackage{graphicx}

\usepackage{xcolor}   
\usepackage[toc,page,titletoc]{appendix}
\usepackage{bm}        
\usepackage{amssymb}   
\usepackage{physics}
\usepackage{tikz}
\usepackage{gensymb}
\usepackage{subfigure}
\usepackage{hyperref}
\usepackage{xcolor}
\usepackage{float}
\usepackage{cleveref}
\usepackage{dsfont}
\usepackage{mathtools}
 \usepackage{amsthm}

\usepackage[normalem]{ulem}

\makeatletter
\DeclareRobustCommand{\element}[1]{\@element#1\@nil}
\def\@element#1#2\@nil{%
  #1%
  \if\relax#2\relax\else\MakeLowercase{#2}\fi}
\makeatother

\begin{document}
\widetext

\title{Quantum Optimal Control of Nuclear Spin Qudecimals in \textsuperscript{87}Sr}

\author{Sivaprasad Omanakuttan}
\email[]{somanakuttan@unm.edu}
\author{Anupam Mitra}
\affiliation{Center for Quantum Information and Control (CQuIC), Department of Physics and Astronomy, University of New Mexico, Albuquerque, New Mexico 87131, USA}
\author{Michael J. Martin }
\affiliation{Materials Physics and Applications Division, Los Alamos National Laboratory, Los Alamos, New Mexico 87544}
\affiliation{Center for Quantum Information and Control (CQuIC), Department of Physics and Astronomy, University of New Mexico, Albuquerque, New Mexico 87131, USA}
\author{Ivan H Deutsch}
\email[]{ideutsch@unm.edu}
\affiliation{Center for Quantum Information and Control (CQuIC), Department of Physics and Astronomy, University of New Mexico, Albuquerque, New Mexico 87131, USA}

\date{\today}
\begin{abstract}
We study the ability to implement unitary maps on states of the $I=9/2$ nuclear spin in \textsuperscript{87}Sr, a $d=10$  dimensional (qudecimal) Hilbert space, using quantum optimal control. Through a combination of nuclear spin-resonance and a tensor AC-Stark shift, by solely modulating the phase of a radio-frequency magnetic field, the system is quantum controllable.  Alkaline earth atoms, such as \textsuperscript{87}Sr, have a very favorable figure-of-merit for such control due to  narrow  intercombination lines and the large hyperfine splitting in the excited states. We numerically study the quantum speed-limit, optimal parameters, and the fidelity of arbitrary state preparation and full SU(10) maps, including the presence of decoherence due to optical pumping induced by the light-shifting laser. We also study the use of robust control to mitigate some dephasing due to inhomogenieties  in the light shift.  We find that with an  rf-Rabi frequency of $\Omega_\text{rf}$ and 0.5\% inhomogeneity  in the the light shift we can prepare an arbitrary Haar-random state in a time  $T={4.5}\pi/\Omega_\text{rf}$  with average fidelity $\langle \mathcal{F}_\psi \rangle =0.9992$, and an arbitrary Haar-random SU(10) map in a time $T=24\pi/\Omega_\text{rf}$ with average fidelity $\langle \mathcal{F}_U \rangle = 0.9923$.

 \end{abstract}
\maketitle

Ultracold ensembles of alkaline-earth atoms trapped in optical lattices or arrays of optical tweezers are a powerful platform for quantum information processing (QIP), including atomic clocks and sensors~\cite{Ludlow2015, campbell2017fermi, norcia2019seconds, covey20192000, young2020half}, simulators of many-body physics~\cite{gorshkov2010two, daley2011quantum, mukherjee2011many,banerjee2013atomic,isaev2016spin, kolkowitz2017spin}, and general purpose quantum computers~\cite{madjarov2020high,daley2011quantum,hayes2007quantum}.  The ability to optically manipulate coherence in single-atoms  via ultranarrow optical resonances on the intercombination lines, together with the ability to create high-fidelity entangling interactions between atoms when they are excited to high-lying Rydberg states~\cite{Saffman2010,Saffman_2016,Browaeys_2016} provides tools that makes this system highly controllable for such applications.  In addition, fermionic species have nuclear spin. As the ground state is a closed shell, there is no electron angular momentum, and the nuclear spin with its weak magnetic moment is highly isolated from the environment.  Such nuclear spins in alkaline-earth atoms are thus natural carriers of quantum information given their long coherence times and our ability to coherently control them with magnetic and optical  fields.  Nuclear spins are also seen as excellent carriers of quantum information in the solid state as demonstrated in pioneering experiments including in NV-centers~\cite{morishita2020} and dopants in silicon~\cite{soltamov2019,morello2018quantum,Godfrin2017,Leuenberger2003}.  

Using magneto-optical fields,~\cite{Lester2021} recently demonstrated the control of qubits encoded in  two nuclear-spin magnetic sublevels levels in  \textsuperscript{87}Sr. The nuclear spin in this atomic species, however, it is not a two-level system; the spin is $I=9/2$ and there are $d=2I+1=10$ nuclear magnetic sublevels.  Such qudits,  here  ``qudecimals," have potential advantage for QIP.  First and foremost, one can encode a $D=d^{n_d}=2^{n_2}$ dimensional Hilbert space associated with $n_2$ qubits in $n_d=n_2/\log_2 d$ qudits.  While only a logarithmic saving, this is meaningful for the qudecimal ($\log_2 d= 3.32$), especially when trapping and control of each atom is at a premium. This savings extends to algorithmic efficiency, in that the number of elementary two-qudit gates necessary to implement a  general unitary map scales as $O(n_d^2 D^2) = O\left( \frac{n_2^2 D^2}{(\log_2 d)^2} \right)$~\cite{muthukrishnan2000multivalued}.  Moreover, qudit architectures can show increased resilience to noise~\cite{cozzolino2019high} and additional routes to quantum error correction~\cite{gottesman1998fault}. For example, one can protect against dephasing errors by encoding a qubit in a nuclear spin qudit~\cite{Li2017}.  In addition, fault-tolerant operation of a quantum computer may be more favorable based on qudit vs. qubit codes~\cite{PhysRevA.83.032310,campbell2014}.

While QIP with qudits has great potential, there are substantial hurdles.  State preparation and readout are more challenging for systems with $d>2$.  Moreover, quantum logic with qudits is more complex.  Universal quantum logic with qubits can be achieved with a set of logic gates that include the unitary-generators of SU(2) on each qubit, plus one entangling gate between qubits pairwise.  In the case of qudits, in addition to the entangling gate, we require unitary-generators of SU($d$) for each subsystem~\cite{muthukrishnan2000multivalued,zhou2003quantum,brennen2005criteria,luo2014universal}. Unlike qubits, the Lie algebra of such gates are not spanned by the native Hamiltonians, and thus implementation of this generating set is not straightforward.  Different approaches have been studied to implement SU($d$) gates~\cite{Moreno2018,neeley2009emulation,Low2020,sawant2020,moro2019}. One approach is to specify an arbitrary SU($d$) unitary matrix through a sequence of so-called Givens rotations acting between pairs of levels~\cite{O'Leary2006}.   In a landmark experiment, the Innsbruck group employed this construction to experimentally demonstrate universal quantum logic with qudits in a trapped ions ion~\cite{ringbauer2021universal}, with performance similar to qubit quantum processors.

An alternative powerful approach to implementing universal quantum logic is to employ the tools of quantum optimal control . In this paradigm, one numerically searches for a time-dependent waveform that achieves the desired SU($d$) unitary map when one has access to a Hamiltonian that makes the system universally ``controllable"~\cite{Merkel2009, jurdjevic1972control, goerz2015optimizing, koch2016controlling,Frey2020 }.  Optimal control is a powerful and flexible approach that does not require specific pairwise Givens rotations, can be high-fidelity, and can be made robust to imperfections such as inhomogenieties through the tools of robust control~\cite{anderson2015accurate, goerz2015optimizing, glaser2015training, koch2016controlling}. In seminal work, the Jessen group used optimal control to demonstrate high-fidelity control of qudits encoded in the hyperfine spin levels of ground-state cesium~\cite{Chaudhury2007, Smith2013}.  This flexible control has found potential application in studies of quantum simulation~\cite{Poggi2020}. 


In this paper we build on this approach to study implementation of SU(10) gates on the nuclear spin of $^{87}$Sr-based on quantum optimal control. A nuclear-spin encoding may have long-term advantages compared to hyperfine states that couple electron and nuclear spins, in its strongly reduced sensitivity to to background magnetic fields and resilience against decoherence driven by photon scattering from optical tweezers or lattices ~\cite{hayes2007quantum, dorscher2018lattice}.  Weak coupling to the environment, of course, comes with increased challenges of weak coupling to control fields.  We will show, nonetheless, that with reasonable experimental parameters one can implement high-fidelity qudecimal logic, with low decoherence.

We consider open loop-control in a Hilbert space with finite dimension $d$, governed by a Hamiltonian $H[\mathbf{c}(t)] = H_0 + \sum_\lambda c_\lambda(t) H_\lambda$ where $\mathbf{c}(t)=\{c_\lambda(t)\}$ is the set of time-dependent classical control waveforms. The system is said to ``controllable" if the set of Hamiltonians, $\{H_0, H_\lambda\}$, are generators of the Lie algebra SU($d$).  Then $\exists \hspace{0.1cm} \mathbf{c}(t)$ \hspace{0.2cm } such that $U[\mathbf{c},T]=\mathcal{T}\left[\exp\left(-i\int_0^T H[\mathbf{c}(t)]dt\right)\right]=U_{\mathrm{tar} }$ for any target unitary matrix $U_{\mathrm{tar} }=\mathrm{SU(d)}$ in this space.  The minimal time $T$ for which this is possible is known as the ``quantum speed limit" (QSL)~\cite{caneva2009optimal} . Additional details of the quantum control protocol used here are described in the supplementary material.

One can achieve quantum controllability of the nuclear spin qudecimal through magneto-optical interactions.  We combine magnetic spin resonance in the presence of an off-resonant laser field as depicted in Fig. 1.  The Hamiltonian acting on the nuclear spin in the $5s^2$ $^1S_0$ ground state takes the form $H=H_{\mathrm{mag}} +H_{\mathrm{LS}}$. Here $H_{\mathrm{mag}} = -\boldsymbol{\mu}\cdot \mathbf{B}(t)$ is the magnetic spin-resonance Hamiltonian, with $\boldsymbol{\mu} = g_I \mu_N \mathbf{I}$ the nuclear magnetic dipole vector operator and $\mathbf{B}(t) = B_\parallel \mathbf{e}_z + B_T \Re\left[(\mathbf{e}_x + i \mathbf{e}_y) \mathbf{e}^{-i\left(\omega_\text{rf} t +\phi(t)\right)}\right]$ the magnetic field consisting of a strong bias defining the quantization axis $\mathbf{e}_z$ and a transversely rotating rf-magnetic field with a time dependent phase $\phi(t)$. Taken alone, the $H_{\mathrm{mag}}$ generates only SU(2) rotations of nuclear spin.  To achieve full SU($d$) control we add a light-shift Hamiltonian due to the AC-Stark effect, $H_{LS}=-\alpha_{zz}(\omega_L) \left| E_0 \right|^2/4$ where $\alpha_{zz}(\omega_L)$ is the $zz$-component of atomic AC-polarizability tensor operator for a laser field at frequency $\omega_L$ linearly-polarized along the quantization axis, $\mathbf{E}_L(t) = \mathbf{e}_z \Re \left(E_0 e^{-i\omega_L t}\right)$. The form of $\alpha_{zz}$ depends on the atomic structure and the detuning of the laser from atomic resonance.  In particular, when the detuning is not large compared to the hyperfine splitting in the excited state, the polarizability has an irreducible rank-2 tensor component $\alpha_{zz} = \alpha^{(2)} I_z^2$ (there also a trivial scalar term proportion to the identity)~\cite{deutsch2010quantum}.  This quadratic spin twist together with the linear Larmor precession yields a set of control Hamiltonians $\{I_x, I_y, I_z^2\}$ sufficient to generate the Lie algebra SU($2I+1$) for an arbitrary spin $I$~\cite{Giorda2003}.  Such control was first demonstrated in the alkali atom cesium, for the hyperfine spin $F=3$ in the electronic ground state, in order to generate nonclassical spin states in the $d=7$ dimensional Hilbert space~\cite{Chaudhury2007}.  

\begin{figure}

\includegraphics[width=0.5\textwidth, height=8cm]{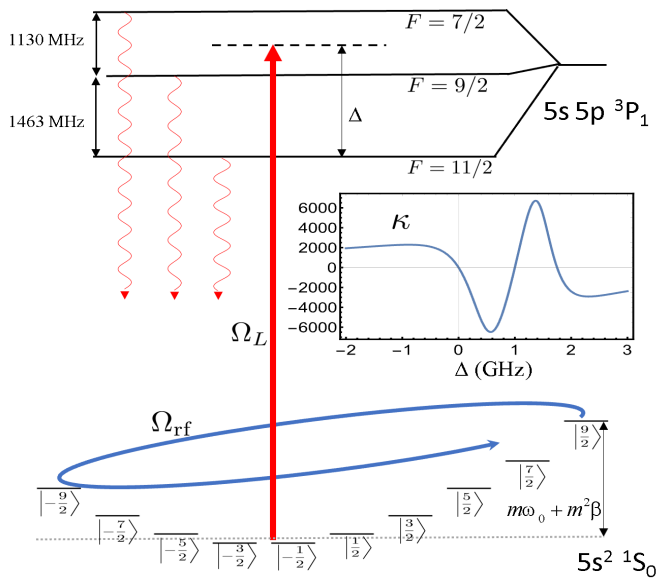}

\caption{  Schematic for magneto-optical control.  The qudecimal is encoded in the ten magnetic sublevels of the nuclear spin, $\ket{-9/2}\rightarrow \ket{9/2}$, in the $5s^2$ $^1S_0$ ground state. Their levels are shifted by a linear Zeeman effect due to a bias magnetic field and a quadratic tensor AC-Stark effect induced by an off-resonant laser beam, polarized along the quantization axis, and detuned $\Delta$ between the hyperfine levels of the $5s5p$ $^3P_1$ intercombination line.  Control of the qudecimal is then achieved with a phase modulated radio-frequency magnetic field, co-rotating at the bare Larmor precession frequency, whose amplitude causes Rabi rotations at frequency $\Omega_{\mathrm{rf}}$. The figure of merit for the control is the ratio of the AC-Stark shift to the photon scattering, $\kappa$ , shown in the inset (see text).}
\label{fig:level_diagram_strontium}
\end{figure}

Importantly, the size of tensor polarizability $\alpha^{(2)}$ depends on the ratio of the excited state hyperfine splitting to the laser detuning~\cite{deutsch2010quantum} , achieving its maximum when these are of the same order.  Thus, to achieve high-fidelity control, one must tune sufficiently close to resonance, while avoiding photon scattering that leads to decoherence. Critically, in alkaline-earth atoms, the first excited $^3P_1$ states have long lifetimes and large hyperfine splittings. This leads to a very favorable figure of merit for optimal control, as measured by the ratio of the characteristic tensor light shift to the photon scattering rate  $\gamma_s$,  $\kappa \equiv \alpha^{(2)} \left| E_0 \right|^2 /4 \gamma_s$.  For example, in \textsuperscript{87}Sr, the hyperfine splitting between the $F=7/2$ and  $F=9/2$ levels in the singly-excited $5s5p \; ^3P_1$ state is $\omega_{\mathrm{HF}}/2\pi=1130$ MHz, while the spontaneous emission linewidth is $\Gamma/2\pi = 7.5$ kHz.   For a scattering rate averaged over all magnetic sublevels \cite{deutsch2010quantum}, we find that when we detune  about halfway between these resonances, we obtain the maximum figure of merit $\kappa =6.8 \times 10^3$ (see Fig. \ref{fig:level_diagram_strontium}).  In contrast, $\kappa =18.6$ for $F=3$ hyperfine spin in the cesium ground state when the laser is tuned halfway between the $F=3$ and $F=4$ hyperfine levels in the excited $6P_{1/2}$ D1-resonance.  This small figure of merit limited the fidelity to around $0.85$ for the arbitrary state preparation.  A factor of $364$ increase in the figure of merit for alkaline earths shows the potential power of this approach to yield high-fidelity quantum optimal control of the nuclear spin qudit.

We consider control of the nuclear spin qudecimal with on-resonance rf fields  on resonance with the Zeeman splitting,  $\Delta E_0 = |g_I| \mu_N B_\parallel$, where $g_I \mu_N/h  = -184$~Hz/Gauss in \textsuperscript{87}Sr~\cite{olschewski1972messung}. In the rotating frame, the control Hamiltonian is
\begin{equation}
H(t)=\Omega_{\mathrm{rf}} \left( \cos[c(t) \pi]I_x+\sin[c(t) \pi] I_y\right)+\beta I_z^2,
\label{eq:Control_Hamiltonian}
\end{equation}
where $\Omega_\text{rf} = -g_I \mu_N B_T $ is the  rf-Rabi frequency and $\beta = \alpha^{(2)} \left|E_0\right|^2/4$ is the strength of the tensor light shift (here and to follow $\hbar=1$). Note, for a rotating rf-field, there is no rotating wave approximation, and this Hamiltonian is valid even when $\Omega_\text{rf} \ge \omega_\text{rf}$. Here the  control waveform is solely the rf-phase $c(t) \equiv \phi(t)/\pi$.   It was proven in \cite{Merkel2008} that varying $c(t)$ is sufficient to achieve universal control the system. 

We consider two classes of quantum control tasks, preparation of a target pure state $\ket{\psi_\text{tar}}$ and implementation of a unitary map $U_\text{tar}$.  Optimal control follows by maximizing the relevant fidelity,
\begin{eqnarray}
\mathcal{F}_{\psi}[\bm{c},T]&=&\left|\bra{\psi_{\text{tar}}}U[\bm{c},T]\ket{\psi_0}\right|^2,\\
\mathcal{F}_U[\bm{c},T]&=&\left|\Tr\left(U^{\dagger}_{\text{tar}}U[\bm{c},T]\right)\right|^2/d^2.
\label{eq:fidelity}
\end{eqnarray} 
This is achieved by discretizing the control waveform and then numerically maximizing the fidelity with gradient ascent.   In a series of works, the Rabitz group showed that the fidelity landscape is favorable for this purpose~\cite{
rabitz2004quantum,Hsieh2008}.  We choose here a piecewise constant parameterization (as in ~\cite{Merkel2008}) and write the control function as a vector $\mathbf{c} = \{c(t_j) | j=1,\dots,n\}$ where $t=j\Delta t$ and $n=T/\Delta t$, parameterizing waveforms that are constant over the duration $\Delta t$.  A minimal choice of $n$ depends on the number of parameters necessary for the control task; for state-maps $n_\text{min} = 2d-2$ and for arbitrary SU($d$) maps $n_\text{min} = d^2-1$.  In practice, we choose $n$ to be a larger than $n_\text{min}$ which improves the fidelity landscape when $T$ is close the the QSL. To numerically optimize $\mathcal{F}$ we use a variation of the well-known GRAPE algorithm~\cite{khaneja2005optimal}. For further details on the choice of parameterization and optimization, see supplemental material.
 
\begin{figure}
\includegraphics[width=0.5\textwidth]{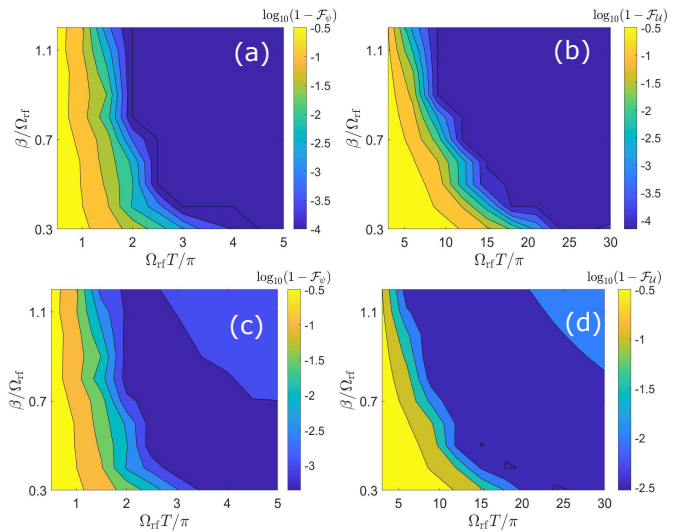}

\caption{Fidelity of objectives found by optimal control as a function of the strength of AC-stark shift, $\beta$, and the total time $T$, in units of the rf-Rabi frequency $\Omega_\text{rf}$. Predictions based on closed-unitary evolution for state-maps (a) and  SU(10) unitary-maps (b) averaged over $120$ Haar-random target states and $10$ Haar-random target SU(10) matrices, respectively.  The control waveforms are piecewise constant, over times $\delta t = T/n$.  For state maps we choose $n=120$ time steps; the unitary maps we take $n=500$. The bottom layer gives the similar figures in the presence of decoherence using the master equation, Eq. (5): state fidelity(c), Eq. (6): and process fidelity (d).}
\label{fig:decohrence_quantum_control}
\end{figure}

For a fixed value of $\Omega_\text{rf}$, the optimal choice of $\beta$ and total time $T$ are found empirically. Figures \ref{fig:decohrence_quantum_control}a(b) show the infidelity, $1-\mathcal{F}$, for state preparation (unitary maps), when averaged over 20 Haar random target vectors (10 random unitary maps).   As expected, when $T \rightarrow \infty$ the infidelity is essentially zero. The QSL is highly dependent on the value of $\beta$.
As expected, the optimal choice is $\beta \approx \Omega_\text{rf}$ as this provides the optimal mixing between Larmor precession and one-axis twisting.  The characteristics of state preparation and unitary maps are  similar in nature. The major difference between these two cases is that unitary mapping requires more time for the simple reason that unitary mapping has $d^2-1$ parameters compared to the $2d-2$ for the state preparation.  The quantum speed limit at $\beta=\Omega_{\mathrm{rf}}$ is $T_* \approx 1.5 \pi/\Omega_{\mathrm{rf}}$ for state preparation and $T_* \approx 8 \pi/\Omega_{\mathrm{rf}}$ for SU(10) unitary maps.
 
In principle, one can achieve arbitrarily high fidelity with increasing $T$.  In practice $T$ is limited by the coherence time of the system.  Here, the coherence time is fundamentally limited by decoherence arising from photon scattering and optical pumping due to the off-resonant light-shift laser.  
We model the effects of decoherence in the state preparation protocols using the Lindblad Master equation \cite{deutsch2010quantum},

\begin{eqnarray}
\frac{d\rho[\bm{c}, t]}{dt} &=&-i \comm{H_\text{eff}[\bm{c}]}{\rho[\bm{c},t]}+ \Gamma\sum_{i}W_q\rho[\bm{c},t] W_{q}^{\dagger}  \nonumber \\ 
&\equiv& \mathcal{L}[\bm{c}]\left[ \rho[\bm{c},t]\right].
\label{eq: evolution of the density matrix}
\end{eqnarray}
where the jump operators for optical pumping between magnetic sublevels describing absorption followed by emission of a $q$-polarized photon are $W_q$,
\begin{equation}
W_q=\sum_{F'}\frac{\Omega/2}{\Delta_{FF'}+i\Gamma/2}(\bm{e}_q^{*}.\bm{D}_{FF'})(\vec{\epsilon}_L.\bm{D}_{FF'}^{\dagger}).
\end{equation}
Here $\bm{D}_{FF'}^{\dagger}$ are the dimensionless dipole raising operators from ground state manifold $F=I$ to the excited state manifold $F'$, as defined in \cite{deutsch2010quantum}. $H_\text{eff}[\mathbf{c}] = H[\mathbf{c}] -i\Gamma \sum_q W_q^\dag W_q$/2 is the non-Hermitian control Hamiltonian, Eq. (\ref{eq: evolution of the density matrix}), now including absorption of the laser light.  

 For gates, we define a $d^2 \times d^2$ superoperator matrix acting on the density matrix.  For the open quantum system, the superoperator describing the evolution of an arbitrary input state is the Completely Positive (CP)-map, $\mathcal{E}[\bm{c},T]=\mathcal{T}\left(\exp\{\int_0^T \mathcal{L}[\bm{c}(t')]\}dt'\right)$, where $\mathcal{L}$ is the Lindbladian superoperator of the master equation,  defined implicitly in Eq. (\ref{eq: evolution of the density matrix}).


 
  We compared the output in the open quantum system dynamics given the  ideal control solution $\mathbf{c}$ found in closed-system optimization. The fidelities for state preparation and full SU(10) maps are, respectively,
\begin{eqnarray}
\mathcal{F}_\psi[\bm{c},T] &=& \Tr{\rho_{\psi_\text{tar}} \rho[\bm{c},T]}, \\
\mathcal{F}_U[\bm{c},T] &=& \left|\Tr{\mathcal{E}_{U_\text{tar}}^{\dagger} \mathcal{E}[\bm{c},T]}\right|/d^2.
\label{eq:effective Fidelity}
\end{eqnarray}
Here $\rho_{\psi_\text{tar}}=\ket{\psi_\text{tar}}\bra{\psi_\text{tar}}$ is the target state and $\rho[\bm{c},T]$ is the solution to the master equation.  $\mathcal{E}_{U_\text{tar}} = U_\text{tar}^{*}\otimes U_\text{tar}$ is the CP-map corresponding to the target unitary gate and $\mathcal{E}[\bm{c},T]$ is the CP-map with decoherence.   Eq. (\ref{eq:effective Fidelity}) is the ``process fidelity," a key quantity of interest in determining the thresholds for fault-tolerant quantum computation \cite{schulte2011}.

Numerical results are given in Fig.~\ref{fig:decohrence_quantum_control} for both state preparation and unitary mapping. In contrast to closed-system control,  Fig.~\ref{fig:decohrence_quantum_control}c and Fig.~\ref{fig:decohrence_quantum_control}d show that there is an island  where the  infidelity is smallest. This reflects the tradeoff between coherent control and decoherence.  There is an optimal total time of evolution $T$ than larger than the QSL but not too large when compared to the optical pumping time.  In addition, the optimal choice of  $\beta$ is now smaller than we found for the closed quantum system, as increased tensor-light shift is accompanied by increased photon scattering. Including decoherence, for the case of state preparation, averaged over $20$ random states, we find the fidelity $\langle \mathcal{F}_\psi \rangle \approx 0.9997$.   Here the island of high fidelity is large, occurring for  $\beta< 1.2$. For the case of unitary mapping the island of lowest infidelity  occurs for  $\beta<1.2$ where the fidelity $\langle \mathcal{F}_U \rangle \approx 0.9970$ which is averaged over $10$ Haar random unitaries.  We emphasize that these qudecimal maps act on a 10-dimension Hilbert space.  Thus a fair comparison of the effective fidelity acting on qubits is $\langle \mathcal{F} \rangle_\text{qubit}= \langle \mathcal{F} \rangle_\text{qudecimal}^{0.3}$, since, in principle, one can encode more than 3 qubits in a qudecimal

Coherence is also limited when there are inhomogenieties arising from uncertainties in the Hamiltonian parameters such as the laser intensity and detuning.  When the decoherence time is longer than than the inhomogeneous dephasing time, one can mitigate this with the numerical tools of robust control~\cite{PhysRevLett.82.2417, PhysRevA.58.2733, anderson2015accurate}.  We consider here an uncertainty in the tensor light shift arising from the thermal velocity of the atoms.  To perform robust control, we replace the control Hamiltonian by $H[\bm{c}] \rightarrow H'[\bm{c},\epsilon]=H[\bm{c}]+\epsilon I_z^2 $, where $\epsilon$ is the variation in $\beta$ around the fiducial value, and define a new objective function as the average fidelity, $\langle \mathcal{F}[\mathbf{c},T] \rangle =\int d\epsilon \; p(\epsilon)\mathcal{F}[\mathbf{c},T, \epsilon]$.  While in principle one can design inhomogeneous control with detailed knowledge of the probability distribution $p(\epsilon)$, in practice, when the standard deviation of the distribution $\delta$ is sufficiently narrow, it is sufficient to simultaneously optimize at two points\cite{anderson2015accurate},  and choose the objective function as 
\begin{equation}
\langle \mathcal{F}[\mathbf{c},T] \rangle = (\mathcal{F}[\mathbf{c},T, \epsilon=+\delta]+\mathcal{F}[\mathbf{c},T, \epsilon=-\delta])/2.
\end{equation}

\begin{figure}

\includegraphics[width=0.5\textwidth]{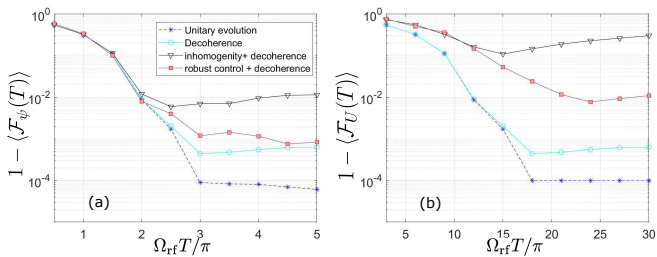}

\caption{ Comparison of infidelity with and without decoherence and robust control to counteract dephasing due to inhomogeneities at the level of $.5\%$ of $\beta$ and $\beta=0.4\Omega_{\mathrm{rf}}$.  (a) state preparation (averaged over $20$ Haar-random target states),  (b) SU(10) mapping (averaged over $10$ Haar-random unitary matrices). Robust control can largely remove dephasing and achieve almost same the infidelity seen due solely to decoherence.}
\label{fig:inhomogenity} 
\end{figure}

The numerical results of robust control are shown in Fig.~\ref{fig:inhomogenity} for $\beta=0.4\Omega_{\mathrm{rf}}$ and an error of  $\delta = .005 \beta$. We see that robust control outperforms the bare waveforms, even in presence of decoherence, but one does not reach the fidelity without any inhomogeneity due to optical pumping occurring over the extended time of the control pulses. For the parameters chosen here, we find that for state preparation one could achieve a fidelity of $\langle \mathcal{F}_\psi \rangle \approx 0.9992$ in a time $T={4.5}\pi/\Omega_\text{rf}$, and for unitary mapping one achieved a fidelity  $\langle \mathcal{F}_U \rangle \approx 0.9923$  in a time $T={24}\pi/\Omega_\text{rf}$. 
Other practical considerations such as the bandwidth needed for rapidly varying waveform may limit the speed of operation (see supplemental material).

We have shown that in the presence of fundamental decoherence and small inhomogeneities, quantum optimal control allows for the realization of high-fidelity arbitrary state maps and SU(10) qudecimal gates acting on nuclear spin in the ground state of \textsuperscript{87}Sr. While we proposed one protocol that leverages the strong tensor light shift induced by a laser tuned near the $^3P_1$ hyperfine manifold, the richness of magneto-optical controls in\textsuperscript{87}Sr provides multiple possible approaches, \textit{e.g.}, by employing the tensor light shift when tuned near the $^3P_0$ clock state.  Quantum optimal control of nuclear spins should find a variety of applications in QIP, including metrological enhancement with qudits~\cite{Noris2012}, quantum simulation~\cite{Poggi2020,Blok2021}, and universal quantum computation~\cite{daley2011quantum}. For the latter additional components are necessary. One must enable readout of all 10 magnetic sublevels though appropriate shelving and fluorescence protocols~\cite{boyd2007nuclear}. Most importantly, we must study the implementation of entangling gates consistent with qudit logic.  Advances in Rydberg-state control for alkaline earth atoms show great promise in this direction~\cite{madjarov2020high}.  Finally, while we have studied here two extremes of the control tasks, state preparation and SU(10) maps, optimal control allows for arbitrary partial isometries to encode a $d'<10$ qudit in the qudecimal.  For example one can encode a qubit in the logical states $\ket{0}=\ket{m_I=9/2}$, $\ket{1}=\ket{m_I=-9/2}$ and potentially protect it from dephasing noise, analogous to a cat-code~\cite{Li2017} or other encodings of a qubit in a large spin that leverages the available interactions and dominant error channels~\cite{Gross2021}.  The flexibility of arbitrary control provides avenues to explore the best approach to encoding and error mitigation. 

\begin{acknowledgements}

This work was supported by the Laboratory Directed Research and Development program of Los Alamos National Laboratory under project numbers 20200015ER,  
and the NSF Quantum Leap Challenge Institutes program, Award No. 2016244. The authors acknowledge fruitful discussions with Jun Ye.
\end{acknowledgements}

\bibliography{QCNS_main}

\begin{thebibliography}{66}
\expandafter\ifx\csname natexlab\endcsname\relax\def\natexlab#1{#1}\fi
\expandafter\ifx\csname bibnamefont\endcsname\relax
  \def\bibnamefont#1{#1}\fi
\expandafter\ifx\csname bibfnamefont\endcsname\relax
  \def\bibfnamefont#1{#1}\fi
\expandafter\ifx\csname citenamefont\endcsname\relax
  \def\citenamefont#1{#1}\fi
\expandafter\ifx\csname url\endcsname\relax
  \def\url#1{\texttt{#1}}\fi
\expandafter\ifx\csname urlprefix\endcsname\relax\def\urlprefix{URL }\fi
\providecommand{\bibinfo}[2]{#2}
\providecommand{\eprint}[2][]{\url{#2}}

\bibitem[{\citenamefont{Ludlow et~al.}(2015)\citenamefont{Ludlow, Boyd, Ye,
  Peik, and Schmidt}}]{Ludlow2015}
\bibinfo{author}{\bibfnamefont{A.~D.} \bibnamefont{Ludlow}},
  \bibinfo{author}{\bibfnamefont{M.~M.} \bibnamefont{Boyd}},
  \bibinfo{author}{\bibfnamefont{J.}~\bibnamefont{Ye}},
  \bibinfo{author}{\bibfnamefont{E.}~\bibnamefont{Peik}}, \bibnamefont{and}
  \bibinfo{author}{\bibfnamefont{P.~O.} \bibnamefont{Schmidt}},
  \bibinfo{journal}{Rev. Mod. Phys.} \textbf{\bibinfo{volume}{87}},
  \bibinfo{pages}{637} (\bibinfo{year}{2015}),
  \urlprefix\url{https://link.aps.org/doi/10.1103/RevModPhys.87.637}.

\bibitem[{\citenamefont{Campbell et~al.}(2017)\citenamefont{Campbell, Hutson,
  Marti, Goban, Oppong, McNally, Sonderhouse, Robinson, Zhang, Bloom
  et~al.}}]{campbell2017fermi}
\bibinfo{author}{\bibfnamefont{S.~L.} \bibnamefont{Campbell}},
  \bibinfo{author}{\bibfnamefont{R.}~\bibnamefont{Hutson}},
  \bibinfo{author}{\bibfnamefont{G.}~\bibnamefont{Marti}},
  \bibinfo{author}{\bibfnamefont{A.}~\bibnamefont{Goban}},
  \bibinfo{author}{\bibfnamefont{N.~D.} \bibnamefont{Oppong}},
  \bibinfo{author}{\bibfnamefont{R.}~\bibnamefont{McNally}},
  \bibinfo{author}{\bibfnamefont{L.}~\bibnamefont{Sonderhouse}},
  \bibinfo{author}{\bibfnamefont{J.}~\bibnamefont{Robinson}},
  \bibinfo{author}{\bibfnamefont{W.}~\bibnamefont{Zhang}},
  \bibinfo{author}{\bibfnamefont{B.}~\bibnamefont{Bloom}},
  \bibnamefont{et~al.}, \bibinfo{journal}{Science}
  \textbf{\bibinfo{volume}{358}}, \bibinfo{pages}{90} (\bibinfo{year}{2017}).

\bibitem[{\citenamefont{Norcia et~al.}(2019)\citenamefont{Norcia, Young,
  Eckner, Oelker, Ye, and Kaufman}}]{norcia2019seconds}
\bibinfo{author}{\bibfnamefont{M.~A.} \bibnamefont{Norcia}},
  \bibinfo{author}{\bibfnamefont{A.~W.} \bibnamefont{Young}},
  \bibinfo{author}{\bibfnamefont{W.~J.} \bibnamefont{Eckner}},
  \bibinfo{author}{\bibfnamefont{E.}~\bibnamefont{Oelker}},
  \bibinfo{author}{\bibfnamefont{J.}~\bibnamefont{Ye}}, \bibnamefont{and}
  \bibinfo{author}{\bibfnamefont{A.~M.} \bibnamefont{Kaufman}},
  \bibinfo{journal}{Science} \textbf{\bibinfo{volume}{366}},
  \bibinfo{pages}{93} (\bibinfo{year}{2019}).

\bibitem[{\citenamefont{Covey et~al.}(2019)\citenamefont{Covey, Madjarov,
  Cooper, and Endres}}]{covey20192000}
\bibinfo{author}{\bibfnamefont{J.~P.} \bibnamefont{Covey}},
  \bibinfo{author}{\bibfnamefont{I.~S.} \bibnamefont{Madjarov}},
  \bibinfo{author}{\bibfnamefont{A.}~\bibnamefont{Cooper}}, \bibnamefont{and}
  \bibinfo{author}{\bibfnamefont{M.}~\bibnamefont{Endres}},
  \bibinfo{journal}{Physical review letters} \textbf{\bibinfo{volume}{122}},
  \bibinfo{pages}{173201} (\bibinfo{year}{2019}).

\bibitem[{\citenamefont{Young et~al.}(2020)\citenamefont{Young, Eckner, Milner,
  Kedar, Norcia, Oelker, Schine, Ye, and Kaufman}}]{young2020half}
\bibinfo{author}{\bibfnamefont{A.~W.} \bibnamefont{Young}},
  \bibinfo{author}{\bibfnamefont{W.~J.} \bibnamefont{Eckner}},
  \bibinfo{author}{\bibfnamefont{W.~R.} \bibnamefont{Milner}},
  \bibinfo{author}{\bibfnamefont{D.}~\bibnamefont{Kedar}},
  \bibinfo{author}{\bibfnamefont{M.~A.} \bibnamefont{Norcia}},
  \bibinfo{author}{\bibfnamefont{E.}~\bibnamefont{Oelker}},
  \bibinfo{author}{\bibfnamefont{N.}~\bibnamefont{Schine}},
  \bibinfo{author}{\bibfnamefont{J.}~\bibnamefont{Ye}}, \bibnamefont{and}
  \bibinfo{author}{\bibfnamefont{A.~M.} \bibnamefont{Kaufman}},
  \bibinfo{journal}{Nature} \textbf{\bibinfo{volume}{588}},
  \bibinfo{pages}{408} (\bibinfo{year}{2020}).

\bibitem[{\citenamefont{Gorshkov et~al.}(2010)\citenamefont{Gorshkov, Hermele,
  Gurarie, Xu, Julienne, Ye, Zoller, Demler, Lukin, and Rey}}]{gorshkov2010two}
\bibinfo{author}{\bibfnamefont{A.~V.} \bibnamefont{Gorshkov}},
  \bibinfo{author}{\bibfnamefont{M.}~\bibnamefont{Hermele}},
  \bibinfo{author}{\bibfnamefont{V.}~\bibnamefont{Gurarie}},
  \bibinfo{author}{\bibfnamefont{C.}~\bibnamefont{Xu}},
  \bibinfo{author}{\bibfnamefont{P.~S.} \bibnamefont{Julienne}},
  \bibinfo{author}{\bibfnamefont{J.}~\bibnamefont{Ye}},
  \bibinfo{author}{\bibfnamefont{P.}~\bibnamefont{Zoller}},
  \bibinfo{author}{\bibfnamefont{E.}~\bibnamefont{Demler}},
  \bibinfo{author}{\bibfnamefont{M.~D.} \bibnamefont{Lukin}}, \bibnamefont{and}
  \bibinfo{author}{\bibfnamefont{A.}~\bibnamefont{Rey}},
  \bibinfo{journal}{Nature physics} \textbf{\bibinfo{volume}{6}},
  \bibinfo{pages}{289} (\bibinfo{year}{2010}).

\bibitem[{\citenamefont{Daley}(2011)}]{daley2011quantum}
\bibinfo{author}{\bibfnamefont{A.~J.} \bibnamefont{Daley}},
  \bibinfo{journal}{Quantum Information Processing}
  \textbf{\bibinfo{volume}{10}}, \bibinfo{pages}{865} (\bibinfo{year}{2011}).

\bibitem[{\citenamefont{Mukherjee et~al.}(2011)\citenamefont{Mukherjee, Millen,
  Nath, Jones, and Pohl}}]{mukherjee2011many}
\bibinfo{author}{\bibfnamefont{R.}~\bibnamefont{Mukherjee}},
  \bibinfo{author}{\bibfnamefont{J.}~\bibnamefont{Millen}},
  \bibinfo{author}{\bibfnamefont{R.}~\bibnamefont{Nath}},
  \bibinfo{author}{\bibfnamefont{M.}~\bibnamefont{Jones}}, \bibnamefont{and}
  \bibinfo{author}{\bibfnamefont{T.}~\bibnamefont{Pohl}},
  \bibinfo{journal}{Journal of Physics B: Atomic, Molecular and Optical
  Physics} \textbf{\bibinfo{volume}{44}}, \bibinfo{pages}{184010}
  (\bibinfo{year}{2011}).

\bibitem[{\citenamefont{Banerjee et~al.}(2013)\citenamefont{Banerjee,
  B{\"o}gli, Dalmonte, Rico, Stebler, Wiese, and Zoller}}]{banerjee2013atomic}
\bibinfo{author}{\bibfnamefont{D.}~\bibnamefont{Banerjee}},
  \bibinfo{author}{\bibfnamefont{M.}~\bibnamefont{B{\"o}gli}},
  \bibinfo{author}{\bibfnamefont{M.}~\bibnamefont{Dalmonte}},
  \bibinfo{author}{\bibfnamefont{E.}~\bibnamefont{Rico}},
  \bibinfo{author}{\bibfnamefont{P.}~\bibnamefont{Stebler}},
  \bibinfo{author}{\bibfnamefont{U.-J.} \bibnamefont{Wiese}}, \bibnamefont{and}
  \bibinfo{author}{\bibfnamefont{P.}~\bibnamefont{Zoller}},
  \bibinfo{journal}{Physical review letters} \textbf{\bibinfo{volume}{110}},
  \bibinfo{pages}{125303} (\bibinfo{year}{2013}).

\bibitem[{\citenamefont{Isaev et~al.}(2016)\citenamefont{Isaev, Schachenmayer,
  and Rey}}]{isaev2016spin}
\bibinfo{author}{\bibfnamefont{L.}~\bibnamefont{Isaev}},
  \bibinfo{author}{\bibfnamefont{J.}~\bibnamefont{Schachenmayer}},
  \bibnamefont{and} \bibinfo{author}{\bibfnamefont{A.}~\bibnamefont{Rey}},
  \bibinfo{journal}{Physical review letters} \textbf{\bibinfo{volume}{117}},
  \bibinfo{pages}{135302} (\bibinfo{year}{2016}).

\bibitem[{\citenamefont{Kolkowitz et~al.}(2017)\citenamefont{Kolkowitz,
  Bromley, Bothwell, Wall, Marti, Koller, Zhang, Rey, and
  Ye}}]{kolkowitz2017spin}
\bibinfo{author}{\bibfnamefont{S.}~\bibnamefont{Kolkowitz}},
  \bibinfo{author}{\bibfnamefont{S.}~\bibnamefont{Bromley}},
  \bibinfo{author}{\bibfnamefont{T.}~\bibnamefont{Bothwell}},
  \bibinfo{author}{\bibfnamefont{M.}~\bibnamefont{Wall}},
  \bibinfo{author}{\bibfnamefont{G.}~\bibnamefont{Marti}},
  \bibinfo{author}{\bibfnamefont{A.}~\bibnamefont{Koller}},
  \bibinfo{author}{\bibfnamefont{X.}~\bibnamefont{Zhang}},
  \bibinfo{author}{\bibfnamefont{A.}~\bibnamefont{Rey}}, \bibnamefont{and}
  \bibinfo{author}{\bibfnamefont{J.}~\bibnamefont{Ye}},
  \bibinfo{journal}{Nature} \textbf{\bibinfo{volume}{542}}, \bibinfo{pages}{66}
  (\bibinfo{year}{2017}).

\bibitem[{\citenamefont{Madjarov et~al.}(2020)\citenamefont{Madjarov, Covey,
  Shaw, Choi, Kale, Cooper, Pichler, Schkolnik, Williams, and
  Endres}}]{madjarov2020high}
\bibinfo{author}{\bibfnamefont{I.~S.} \bibnamefont{Madjarov}},
  \bibinfo{author}{\bibfnamefont{J.~P.} \bibnamefont{Covey}},
  \bibinfo{author}{\bibfnamefont{A.~L.} \bibnamefont{Shaw}},
  \bibinfo{author}{\bibfnamefont{J.}~\bibnamefont{Choi}},
  \bibinfo{author}{\bibfnamefont{A.}~\bibnamefont{Kale}},
  \bibinfo{author}{\bibfnamefont{A.}~\bibnamefont{Cooper}},
  \bibinfo{author}{\bibfnamefont{H.}~\bibnamefont{Pichler}},
  \bibinfo{author}{\bibfnamefont{V.}~\bibnamefont{Schkolnik}},
  \bibinfo{author}{\bibfnamefont{J.~R.} \bibnamefont{Williams}},
  \bibnamefont{and} \bibinfo{author}{\bibfnamefont{M.}~\bibnamefont{Endres}},
  \bibinfo{journal}{Nature Physics} \textbf{\bibinfo{volume}{16}},
  \bibinfo{pages}{857} (\bibinfo{year}{2020}).

\bibitem[{\citenamefont{Hayes et~al.}(2007)\citenamefont{Hayes, Julienne, and
  Deutsch}}]{hayes2007quantum}
\bibinfo{author}{\bibfnamefont{D.}~\bibnamefont{Hayes}},
  \bibinfo{author}{\bibfnamefont{P.~S.} \bibnamefont{Julienne}},
  \bibnamefont{and} \bibinfo{author}{\bibfnamefont{I.~H.}
  \bibnamefont{Deutsch}}, \bibinfo{journal}{Physical Review Letters}
  \textbf{\bibinfo{volume}{98}}, \bibinfo{pages}{070501}
  (\bibinfo{year}{2007}).

\bibitem[{\citenamefont{Saffman et~al.}(2010)\citenamefont{Saffman, Walker, and
  M\o{}lmer}}]{Saffman2010}
\bibinfo{author}{\bibfnamefont{M.}~\bibnamefont{Saffman}},
  \bibinfo{author}{\bibfnamefont{T.~G.} \bibnamefont{Walker}},
  \bibnamefont{and}
  \bibinfo{author}{\bibfnamefont{K.}~\bibnamefont{M\o{}lmer}},
  \bibinfo{journal}{Rev. Mod. Phys.} \textbf{\bibinfo{volume}{82}},
  \bibinfo{pages}{2313} (\bibinfo{year}{2010}),
  \urlprefix\url{https://link.aps.org/doi/10.1103/RevModPhys.82.2313}.

\bibitem[{\citenamefont{Saffman}(2016)}]{Saffman_2016}
\bibinfo{author}{\bibfnamefont{M.}~\bibnamefont{Saffman}},
  \bibinfo{journal}{Journal of Physics B: Atomic, Molecular and Optical
  Physics} \textbf{\bibinfo{volume}{49}}, \bibinfo{pages}{202001}
  (\bibinfo{year}{2016}),
  \urlprefix\url{https://doi.org/10.1088/0953-4075/49/20/202001}.

\bibitem[{\citenamefont{Browaeys et~al.}(2016)\citenamefont{Browaeys, Barredo,
  and Lahaye}}]{Browaeys_2016}
\bibinfo{author}{\bibfnamefont{A.}~\bibnamefont{Browaeys}},
  \bibinfo{author}{\bibfnamefont{D.}~\bibnamefont{Barredo}}, \bibnamefont{and}
  \bibinfo{author}{\bibfnamefont{T.}~\bibnamefont{Lahaye}},
  \bibinfo{journal}{Journal of Physics B: Atomic, Molecular and Optical
  Physics} \textbf{\bibinfo{volume}{49}}, \bibinfo{pages}{152001}
  (\bibinfo{year}{2016}),
  \urlprefix\url{https://doi.org/10.1088/0953-4075/49/15/152001}.

\bibitem[{\citenamefont{Morishita et~al.}(2020)\citenamefont{Morishita,
  Kobayashi, Fujiwara, Kato, Makino, Yamasaki, and Mizuochi}}]{morishita2020}
\bibinfo{author}{\bibfnamefont{H.}~\bibnamefont{Morishita}},
  \bibinfo{author}{\bibfnamefont{S.}~\bibnamefont{Kobayashi}},
  \bibinfo{author}{\bibfnamefont{M.}~\bibnamefont{Fujiwara}},
  \bibinfo{author}{\bibfnamefont{H.}~\bibnamefont{Kato}},
  \bibinfo{author}{\bibfnamefont{T.}~\bibnamefont{Makino}},
  \bibinfo{author}{\bibfnamefont{S.}~\bibnamefont{Yamasaki}}, \bibnamefont{and}
  \bibinfo{author}{\bibfnamefont{N.}~\bibnamefont{Mizuochi}},
  \bibinfo{journal}{Scientific reports} \textbf{\bibinfo{volume}{10}},
  \bibinfo{pages}{1} (\bibinfo{year}{2020}).

\bibitem[{\citenamefont{Soltamov et~al.}(2019)\citenamefont{Soltamov, Kasper,
  Poshakinskiy, Anisimov, Mokhov, Sperlich, Tarasenko, Baranov, Astakhov, and
  Dyakonov}}]{soltamov2019}
\bibinfo{author}{\bibfnamefont{V.}~\bibnamefont{Soltamov}},
  \bibinfo{author}{\bibfnamefont{C.}~\bibnamefont{Kasper}},
  \bibinfo{author}{\bibfnamefont{A.}~\bibnamefont{Poshakinskiy}},
  \bibinfo{author}{\bibfnamefont{A.}~\bibnamefont{Anisimov}},
  \bibinfo{author}{\bibfnamefont{E.}~\bibnamefont{Mokhov}},
  \bibinfo{author}{\bibfnamefont{A.}~\bibnamefont{Sperlich}},
  \bibinfo{author}{\bibfnamefont{S.}~\bibnamefont{Tarasenko}},
  \bibinfo{author}{\bibfnamefont{P.}~\bibnamefont{Baranov}},
  \bibinfo{author}{\bibfnamefont{G.}~\bibnamefont{Astakhov}}, \bibnamefont{and}
  \bibinfo{author}{\bibfnamefont{V.}~\bibnamefont{Dyakonov}},
  \bibinfo{journal}{Nature communications} \textbf{\bibinfo{volume}{10}},
  \bibinfo{pages}{1} (\bibinfo{year}{2019}).

\bibitem[{\citenamefont{Morello}(2018)}]{morello2018quantum}
\bibinfo{author}{\bibfnamefont{A.}~\bibnamefont{Morello}},
  \bibinfo{journal}{Nature nanotechnology} \textbf{\bibinfo{volume}{13}},
  \bibinfo{pages}{9} (\bibinfo{year}{2018}).

\bibitem[{\citenamefont{Godfrin et~al.}(2017)\citenamefont{Godfrin, Ferhat,
  Ballou, Klyatskaya, Ruben, Wernsdorfer, and Balestro}}]{Godfrin2017}
\bibinfo{author}{\bibfnamefont{C.}~\bibnamefont{Godfrin}},
  \bibinfo{author}{\bibfnamefont{A.}~\bibnamefont{Ferhat}},
  \bibinfo{author}{\bibfnamefont{R.}~\bibnamefont{Ballou}},
  \bibinfo{author}{\bibfnamefont{S.}~\bibnamefont{Klyatskaya}},
  \bibinfo{author}{\bibfnamefont{M.}~\bibnamefont{Ruben}},
  \bibinfo{author}{\bibfnamefont{W.}~\bibnamefont{Wernsdorfer}},
  \bibnamefont{and} \bibinfo{author}{\bibfnamefont{F.}~\bibnamefont{Balestro}},
  \bibinfo{journal}{Phys. Rev. Lett.} \textbf{\bibinfo{volume}{119}},
  \bibinfo{pages}{187702} (\bibinfo{year}{2017}),
  \urlprefix\url{https://link.aps.org/doi/10.1103/PhysRevLett.119.187702}.

\bibitem[{\citenamefont{Leuenberger and Loss}(2003)}]{Leuenberger2003}
\bibinfo{author}{\bibfnamefont{M.~N.} \bibnamefont{Leuenberger}}
  \bibnamefont{and} \bibinfo{author}{\bibfnamefont{D.}~\bibnamefont{Loss}},
  \bibinfo{journal}{Phys. Rev. B} \textbf{\bibinfo{volume}{68}},
  \bibinfo{pages}{165317} (\bibinfo{year}{2003}),
  \urlprefix\url{https://link.aps.org/doi/10.1103/PhysRevB.68.165317}.

\bibitem[{\citenamefont{Lester et~al.}(2021)\citenamefont{Lester, Kotru,
  McDonald, Notermans, Cassella, Ryou, Kondov, Peng, Battaglino, Lauigan
  et~al.}}]{Lester2021}
\bibinfo{author}{\bibfnamefont{B.}~\bibnamefont{Lester}},
  \bibinfo{author}{\bibfnamefont{K.}~\bibnamefont{Kotru}},
  \bibinfo{author}{\bibfnamefont{M.~P.} \bibnamefont{McDonald}},
  \bibinfo{author}{\bibfnamefont{R.~P.} \bibnamefont{Notermans}},
  \bibinfo{author}{\bibfnamefont{K.}~\bibnamefont{Cassella}},
  \bibinfo{author}{\bibfnamefont{A.}~\bibnamefont{Ryou}},
  \bibinfo{author}{\bibfnamefont{S.}~\bibnamefont{Kondov}},
  \bibinfo{author}{\bibfnamefont{L.}~\bibnamefont{Peng}},
  \bibinfo{author}{\bibfnamefont{P.}~\bibnamefont{Battaglino}},
  \bibinfo{author}{\bibfnamefont{J.}~\bibnamefont{Lauigan}},
  \bibnamefont{et~al.} (\bibinfo{year}{2021}), \bibinfo{note}{aPS March Meeting
  2021}, \urlprefix\url{http://meetings.aps.org/Meeting/MAR21/Session/M31.1}.

\bibitem[{\citenamefont{Muthukrishnan and
  Stroud~Jr}(2000)}]{muthukrishnan2000multivalued}
\bibinfo{author}{\bibfnamefont{A.}~\bibnamefont{Muthukrishnan}}
  \bibnamefont{and} \bibinfo{author}{\bibfnamefont{C.~R.}
  \bibnamefont{Stroud~Jr}}, \bibinfo{journal}{Physical review A}
  \textbf{\bibinfo{volume}{62}}, \bibinfo{pages}{052309}
  (\bibinfo{year}{2000}).

\bibitem[{\citenamefont{Cozzolino et~al.}(2019)\citenamefont{Cozzolino, Da~Lio,
  Bacco, and Oxenl{\o}we}}]{cozzolino2019high}
\bibinfo{author}{\bibfnamefont{D.}~\bibnamefont{Cozzolino}},
  \bibinfo{author}{\bibfnamefont{B.}~\bibnamefont{Da~Lio}},
  \bibinfo{author}{\bibfnamefont{D.}~\bibnamefont{Bacco}}, \bibnamefont{and}
  \bibinfo{author}{\bibfnamefont{L.~K.} \bibnamefont{Oxenl{\o}we}},
  \bibinfo{journal}{Advanced Quantum Technologies}
  \textbf{\bibinfo{volume}{2}}, \bibinfo{pages}{1900038}
  (\bibinfo{year}{2019}).

\bibitem[{\citenamefont{Gottesman}(1998)}]{gottesman1998fault}
\bibinfo{author}{\bibfnamefont{D.}~\bibnamefont{Gottesman}}, in
  \emph{\bibinfo{booktitle}{NASA International Conference on Quantum Computing
  and Quantum Communications}} (\bibinfo{organization}{Springer},
  \bibinfo{year}{1998}), pp. \bibinfo{pages}{302--313}.

\bibitem[{\citenamefont{Li et~al.}(2017)\citenamefont{Li, Zou, Albert,
  Muralidharan, Girvin, and Jiang}}]{Li2017}
\bibinfo{author}{\bibfnamefont{L.}~\bibnamefont{Li}},
  \bibinfo{author}{\bibfnamefont{C.-L.} \bibnamefont{Zou}},
  \bibinfo{author}{\bibfnamefont{V.~V.} \bibnamefont{Albert}},
  \bibinfo{author}{\bibfnamefont{S.}~\bibnamefont{Muralidharan}},
  \bibinfo{author}{\bibfnamefont{S.~M.} \bibnamefont{Girvin}},
  \bibnamefont{and} \bibinfo{author}{\bibfnamefont{L.}~\bibnamefont{Jiang}},
  \bibinfo{journal}{Phys. Rev. Lett.} \textbf{\bibinfo{volume}{119}},
  \bibinfo{pages}{030502} (\bibinfo{year}{2017}),
  \urlprefix\url{https://link.aps.org/doi/10.1103/PhysRevLett.119.030502}.

\bibitem[{\citenamefont{van Dam and Howard}(2011)}]{PhysRevA.83.032310}
\bibinfo{author}{\bibfnamefont{W.}~\bibnamefont{van Dam}} \bibnamefont{and}
  \bibinfo{author}{\bibfnamefont{M.}~\bibnamefont{Howard}},
  \bibinfo{journal}{Phys. Rev. A} \textbf{\bibinfo{volume}{83}},
  \bibinfo{pages}{032310} (\bibinfo{year}{2011}),
  \urlprefix\url{https://link.aps.org/doi/10.1103/PhysRevA.83.032310}.

\bibitem[{\citenamefont{Campbell}(2014)}]{campbell2014}
\bibinfo{author}{\bibfnamefont{E.~T.} \bibnamefont{Campbell}},
  \bibinfo{journal}{Phys. Rev. Lett.} \textbf{\bibinfo{volume}{113}},
  \bibinfo{pages}{230501} (\bibinfo{year}{2014}),
  \urlprefix\url{https://link.aps.org/doi/10.1103/PhysRevLett.113.230501}.

\bibitem[{\citenamefont{Zhou et~al.}(2003)\citenamefont{Zhou, Zeng, Xu, and
  Sun}}]{zhou2003quantum}
\bibinfo{author}{\bibfnamefont{D.}~\bibnamefont{Zhou}},
  \bibinfo{author}{\bibfnamefont{B.}~\bibnamefont{Zeng}},
  \bibinfo{author}{\bibfnamefont{Z.}~\bibnamefont{Xu}}, \bibnamefont{and}
  \bibinfo{author}{\bibfnamefont{C.}~\bibnamefont{Sun}},
  \bibinfo{journal}{Physical Review A} \textbf{\bibinfo{volume}{68}},
  \bibinfo{pages}{062303} (\bibinfo{year}{2003}).

\bibitem[{\citenamefont{Brennen et~al.}(2005)\citenamefont{Brennen, O’Leary,
  and Bullock}}]{brennen2005criteria}
\bibinfo{author}{\bibfnamefont{G.~K.} \bibnamefont{Brennen}},
  \bibinfo{author}{\bibfnamefont{D.~P.} \bibnamefont{O’Leary}},
  \bibnamefont{and} \bibinfo{author}{\bibfnamefont{S.~S.}
  \bibnamefont{Bullock}}, \bibinfo{journal}{Physical Review A}
  \textbf{\bibinfo{volume}{71}}, \bibinfo{pages}{052318}
  (\bibinfo{year}{2005}).

\bibitem[{\citenamefont{Luo and Wang}(2014)}]{luo2014universal}
\bibinfo{author}{\bibfnamefont{M.}~\bibnamefont{Luo}} \bibnamefont{and}
  \bibinfo{author}{\bibfnamefont{X.}~\bibnamefont{Wang}},
  \bibinfo{journal}{Science China Physics, Mechanics \& Astronomy}
  \textbf{\bibinfo{volume}{57}}, \bibinfo{pages}{1712} (\bibinfo{year}{2014}).

\bibitem[{\citenamefont{Moreno-Pineda et~al.}(2018)\citenamefont{Moreno-Pineda,
  Godfrin, Balestro, Wernsdorfer, and Ruben}}]{Moreno2018}
\bibinfo{author}{\bibfnamefont{E.}~\bibnamefont{Moreno-Pineda}},
  \bibinfo{author}{\bibfnamefont{C.}~\bibnamefont{Godfrin}},
  \bibinfo{author}{\bibfnamefont{F.}~\bibnamefont{Balestro}},
  \bibinfo{author}{\bibfnamefont{W.}~\bibnamefont{Wernsdorfer}},
  \bibnamefont{and} \bibinfo{author}{\bibfnamefont{M.}~\bibnamefont{Ruben}},
  \bibinfo{journal}{Chem. Soc. Rev.} \textbf{\bibinfo{volume}{47}},
  \bibinfo{pages}{501} (\bibinfo{year}{2018}),
  \urlprefix\url{http://dx.doi.org/10.1039/C5CS00933B}.

\bibitem[{\citenamefont{Neeley et~al.}(2009)\citenamefont{Neeley, Ansmann,
  Bialczak, Hofheinz, Lucero, O'Connell, Sank, Wang, Wenner, Cleland
  et~al.}}]{neeley2009emulation}
\bibinfo{author}{\bibfnamefont{M.}~\bibnamefont{Neeley}},
  \bibinfo{author}{\bibfnamefont{M.}~\bibnamefont{Ansmann}},
  \bibinfo{author}{\bibfnamefont{R.~C.} \bibnamefont{Bialczak}},
  \bibinfo{author}{\bibfnamefont{M.}~\bibnamefont{Hofheinz}},
  \bibinfo{author}{\bibfnamefont{E.}~\bibnamefont{Lucero}},
  \bibinfo{author}{\bibfnamefont{A.~D.} \bibnamefont{O'Connell}},
  \bibinfo{author}{\bibfnamefont{D.}~\bibnamefont{Sank}},
  \bibinfo{author}{\bibfnamefont{H.}~\bibnamefont{Wang}},
  \bibinfo{author}{\bibfnamefont{J.}~\bibnamefont{Wenner}},
  \bibinfo{author}{\bibfnamefont{A.~N.} \bibnamefont{Cleland}},
  \bibnamefont{et~al.}, \bibinfo{journal}{Science}
  \textbf{\bibinfo{volume}{325}}, \bibinfo{pages}{722} (\bibinfo{year}{2009}).

\bibitem[{\citenamefont{Low et~al.}(2020)\citenamefont{Low, White, Cox, Day,
  and Senko}}]{Low2020}
\bibinfo{author}{\bibfnamefont{P.~J.} \bibnamefont{Low}},
  \bibinfo{author}{\bibfnamefont{B.~M.} \bibnamefont{White}},
  \bibinfo{author}{\bibfnamefont{A.~A.} \bibnamefont{Cox}},
  \bibinfo{author}{\bibfnamefont{M.~L.} \bibnamefont{Day}}, \bibnamefont{and}
  \bibinfo{author}{\bibfnamefont{C.}~\bibnamefont{Senko}},
  \bibinfo{journal}{Phys. Rev. Research} \textbf{\bibinfo{volume}{2}},
  \bibinfo{pages}{033128} (\bibinfo{year}{2020}),
  \urlprefix\url{https://link.aps.org/doi/10.1103/PhysRevResearch.2.033128}.

\bibitem[{\citenamefont{Sawant et~al.}(2020)\citenamefont{Sawant, Blackmore,
  Gregory, Mur-Petit, Jaksch, Aldegunde, Hutson, Tarbutt, and
  Cornish}}]{sawant2020}
\bibinfo{author}{\bibfnamefont{R.}~\bibnamefont{Sawant}},
  \bibinfo{author}{\bibfnamefont{J.~A.} \bibnamefont{Blackmore}},
  \bibinfo{author}{\bibfnamefont{P.~D.} \bibnamefont{Gregory}},
  \bibinfo{author}{\bibfnamefont{J.}~\bibnamefont{Mur-Petit}},
  \bibinfo{author}{\bibfnamefont{D.}~\bibnamefont{Jaksch}},
  \bibinfo{author}{\bibfnamefont{J.}~\bibnamefont{Aldegunde}},
  \bibinfo{author}{\bibfnamefont{J.~M.} \bibnamefont{Hutson}},
  \bibinfo{author}{\bibfnamefont{M.}~\bibnamefont{Tarbutt}}, \bibnamefont{and}
  \bibinfo{author}{\bibfnamefont{S.~L.} \bibnamefont{Cornish}},
  \bibinfo{journal}{New Journal of Physics} \textbf{\bibinfo{volume}{22}},
  \bibinfo{pages}{013027} (\bibinfo{year}{2020}).

\bibitem[{\citenamefont{Moro et~al.}(2019)\citenamefont{Moro, Fielding,
  Turyanska, and Patan{\`e}}}]{moro2019}
\bibinfo{author}{\bibfnamefont{F.}~\bibnamefont{Moro}},
  \bibinfo{author}{\bibfnamefont{A.~J.} \bibnamefont{Fielding}},
  \bibinfo{author}{\bibfnamefont{L.}~\bibnamefont{Turyanska}},
  \bibnamefont{and}
  \bibinfo{author}{\bibfnamefont{A.}~\bibnamefont{Patan{\`e}}},
  \bibinfo{journal}{Advanced Quantum Technologies}
  \textbf{\bibinfo{volume}{2}}, \bibinfo{pages}{1900017}
  (\bibinfo{year}{2019}).

\bibitem[{\citenamefont{O'Leary et~al.}(2006)\citenamefont{O'Leary, Brennen,
  and Bullock}}]{O'Leary2006}
\bibinfo{author}{\bibfnamefont{D.~P.} \bibnamefont{O'Leary}},
  \bibinfo{author}{\bibfnamefont{G.~K.} \bibnamefont{Brennen}},
  \bibnamefont{and} \bibinfo{author}{\bibfnamefont{S.~S.}
  \bibnamefont{Bullock}}, \bibinfo{journal}{Phys. Rev. A}
  \textbf{\bibinfo{volume}{74}}, \bibinfo{pages}{032334}
  (\bibinfo{year}{2006}),
  \urlprefix\url{https://link.aps.org/doi/10.1103/PhysRevA.74.032334}.

\bibitem[{\citenamefont{Ringbauer et~al.}(2021)\citenamefont{Ringbauer, Meth,
  Postler, Stricker, Blatt, Schindler, and Monz}}]{ringbauer2021universal}
\bibinfo{author}{\bibfnamefont{M.}~\bibnamefont{Ringbauer}},
  \bibinfo{author}{\bibfnamefont{M.}~\bibnamefont{Meth}},
  \bibinfo{author}{\bibfnamefont{L.}~\bibnamefont{Postler}},
  \bibinfo{author}{\bibfnamefont{R.}~\bibnamefont{Stricker}},
  \bibinfo{author}{\bibfnamefont{R.}~\bibnamefont{Blatt}},
  \bibinfo{author}{\bibfnamefont{P.}~\bibnamefont{Schindler}},
  \bibnamefont{and} \bibinfo{author}{\bibfnamefont{T.}~\bibnamefont{Monz}},
  \bibinfo{journal}{arXiv preprint arXiv:2109.06903}  (\bibinfo{year}{2021}).

\bibitem[{\citenamefont{Merkel et~al.}(2009)\citenamefont{Merkel, Brennen,
  Jessen, and Deutsch}}]{Merkel2009}
\bibinfo{author}{\bibfnamefont{S.~T.} \bibnamefont{Merkel}},
  \bibinfo{author}{\bibfnamefont{G.}~\bibnamefont{Brennen}},
  \bibinfo{author}{\bibfnamefont{P.~S.} \bibnamefont{Jessen}},
  \bibnamefont{and} \bibinfo{author}{\bibfnamefont{I.~H.}
  \bibnamefont{Deutsch}}, \bibinfo{journal}{Phys. Rev. A}
  \textbf{\bibinfo{volume}{80}}, \bibinfo{pages}{023424}
  (\bibinfo{year}{2009}),
  \urlprefix\url{https://link.aps.org/doi/10.1103/PhysRevA.80.023424}.

\bibitem[{\citenamefont{Jurdjevic and Sussmann}(1972)}]{jurdjevic1972control}
\bibinfo{author}{\bibfnamefont{V.}~\bibnamefont{Jurdjevic}} \bibnamefont{and}
  \bibinfo{author}{\bibfnamefont{H.~J.} \bibnamefont{Sussmann}},
  \bibinfo{journal}{Journal of Differential equations}
  \textbf{\bibinfo{volume}{12}}, \bibinfo{pages}{313} (\bibinfo{year}{1972}).

\bibitem[{\citenamefont{Goerz}(2015)}]{goerz2015optimizing}
\bibinfo{author}{\bibfnamefont{M.~H.} \bibnamefont{Goerz}}, Ph.D. thesis
  (\bibinfo{year}{2015}).

\bibitem[{\citenamefont{Koch}(2016)}]{koch2016controlling}
\bibinfo{author}{\bibfnamefont{C.~P.} \bibnamefont{Koch}},
  \bibinfo{journal}{Journal of Physics: Condensed Matter}
  \textbf{\bibinfo{volume}{28}}, \bibinfo{pages}{213001}
  (\bibinfo{year}{2016}).

\bibitem[{\citenamefont{Anderson et~al.}(2015)\citenamefont{Anderson,
  Sosa-Martinez, Riofr{\'\i}o, Deutsch, and Jessen}}]{anderson2015accurate}
\bibinfo{author}{\bibfnamefont{B.}~\bibnamefont{Anderson}},
  \bibinfo{author}{\bibfnamefont{H.}~\bibnamefont{Sosa-Martinez}},
  \bibinfo{author}{\bibfnamefont{C.}~\bibnamefont{Riofr{\'\i}o}},
  \bibinfo{author}{\bibfnamefont{I.~H.} \bibnamefont{Deutsch}},
  \bibnamefont{and} \bibinfo{author}{\bibfnamefont{P.~S.}
  \bibnamefont{Jessen}}, \bibinfo{journal}{Physical review letters}
  \textbf{\bibinfo{volume}{114}}, \bibinfo{pages}{240401}
  (\bibinfo{year}{2015}).

\bibitem[{\citenamefont{Glaser et~al.}(2015)\citenamefont{Glaser, Boscain,
  Calarco, Koch, K{\"o}ckenberger, Kosloff, Kuprov, Luy, Schirmer,
  Schulte-Herbr{\"u}ggen et~al.}}]{glaser2015training}
\bibinfo{author}{\bibfnamefont{S.~J.} \bibnamefont{Glaser}},
  \bibinfo{author}{\bibfnamefont{U.}~\bibnamefont{Boscain}},
  \bibinfo{author}{\bibfnamefont{T.}~\bibnamefont{Calarco}},
  \bibinfo{author}{\bibfnamefont{C.~P.} \bibnamefont{Koch}},
  \bibinfo{author}{\bibfnamefont{W.}~\bibnamefont{K{\"o}ckenberger}},
  \bibinfo{author}{\bibfnamefont{R.}~\bibnamefont{Kosloff}},
  \bibinfo{author}{\bibfnamefont{I.}~\bibnamefont{Kuprov}},
  \bibinfo{author}{\bibfnamefont{B.}~\bibnamefont{Luy}},
  \bibinfo{author}{\bibfnamefont{S.}~\bibnamefont{Schirmer}},
  \bibinfo{author}{\bibfnamefont{T.}~\bibnamefont{Schulte-Herbr{\"u}ggen}},
  \bibnamefont{et~al.}, \bibinfo{journal}{The European Physical Journal D}
  \textbf{\bibinfo{volume}{69}}, \bibinfo{pages}{1} (\bibinfo{year}{2015}).

\bibitem[{\citenamefont{Chaudhury et~al.}(2007)\citenamefont{Chaudhury, Merkel,
  Herr, Silberfarb, Deutsch, and Jessen}}]{Chaudhury2007}
\bibinfo{author}{\bibfnamefont{S.}~\bibnamefont{Chaudhury}},
  \bibinfo{author}{\bibfnamefont{S.}~\bibnamefont{Merkel}},
  \bibinfo{author}{\bibfnamefont{T.}~\bibnamefont{Herr}},
  \bibinfo{author}{\bibfnamefont{A.}~\bibnamefont{Silberfarb}},
  \bibinfo{author}{\bibfnamefont{I.~H.} \bibnamefont{Deutsch}},
  \bibnamefont{and} \bibinfo{author}{\bibfnamefont{P.~S.}
  \bibnamefont{Jessen}}, \bibinfo{journal}{Phys. Rev. Lett.}
  \textbf{\bibinfo{volume}{99}}, \bibinfo{pages}{163002}
  (\bibinfo{year}{2007}),
  \urlprefix\url{https://link.aps.org/doi/10.1103/PhysRevLett.99.163002}.

\bibitem[{\citenamefont{Smith et~al.}(2013)\citenamefont{Smith, Anderson,
  Sosa-Martinez, Riofr\'{\i}o, Deutsch, and Jessen}}]{Smith2013}
\bibinfo{author}{\bibfnamefont{A.}~\bibnamefont{Smith}},
  \bibinfo{author}{\bibfnamefont{B.~E.} \bibnamefont{Anderson}},
  \bibinfo{author}{\bibfnamefont{H.}~\bibnamefont{Sosa-Martinez}},
  \bibinfo{author}{\bibfnamefont{C.~A.} \bibnamefont{Riofr\'{\i}o}},
  \bibinfo{author}{\bibfnamefont{I.~H.} \bibnamefont{Deutsch}},
  \bibnamefont{and} \bibinfo{author}{\bibfnamefont{P.~S.}
  \bibnamefont{Jessen}}, \bibinfo{journal}{Phys. Rev. Lett.}
  \textbf{\bibinfo{volume}{111}}, \bibinfo{pages}{170502}
  (\bibinfo{year}{2013}),
  \urlprefix\url{https://link.aps.org/doi/10.1103/PhysRevLett.111.170502}.

\bibitem[{\citenamefont{Poggi et~al.}(2020)\citenamefont{Poggi, Lysne, Kuper,
  Deutsch, and Jessen}}]{Poggi2020}
\bibinfo{author}{\bibfnamefont{P.~M.} \bibnamefont{Poggi}},
  \bibinfo{author}{\bibfnamefont{N.~K.} \bibnamefont{Lysne}},
  \bibinfo{author}{\bibfnamefont{K.~W.} \bibnamefont{Kuper}},
  \bibinfo{author}{\bibfnamefont{I.~H.} \bibnamefont{Deutsch}},
  \bibnamefont{and} \bibinfo{author}{\bibfnamefont{P.~S.}
  \bibnamefont{Jessen}}, \bibinfo{journal}{PRX Quantum}
  \textbf{\bibinfo{volume}{1}}, \bibinfo{pages}{020308} (\bibinfo{year}{2020}),
  \urlprefix\url{https://link.aps.org/doi/10.1103/PRXQuantum.1.020308}.

\bibitem[{\citenamefont{D{\"o}rscher et~al.}(2018)\citenamefont{D{\"o}rscher,
  Schwarz, Al-Masoudi, Falke, Sterr, and Lisdat}}]{dorscher2018lattice}
\bibinfo{author}{\bibfnamefont{S.}~\bibnamefont{D{\"o}rscher}},
  \bibinfo{author}{\bibfnamefont{R.}~\bibnamefont{Schwarz}},
  \bibinfo{author}{\bibfnamefont{A.}~\bibnamefont{Al-Masoudi}},
  \bibinfo{author}{\bibfnamefont{S.}~\bibnamefont{Falke}},
  \bibinfo{author}{\bibfnamefont{U.}~\bibnamefont{Sterr}}, \bibnamefont{and}
  \bibinfo{author}{\bibfnamefont{C.}~\bibnamefont{Lisdat}},
  \bibinfo{journal}{Physical Review A} \textbf{\bibinfo{volume}{97}},
  \bibinfo{pages}{063419} (\bibinfo{year}{2018}).

\bibitem[{\citenamefont{Caneva et~al.}(2009)\citenamefont{Caneva, Murphy,
  Calarco, Fazio, Montangero, Giovannetti, and Santoro}}]{caneva2009optimal}
\bibinfo{author}{\bibfnamefont{T.}~\bibnamefont{Caneva}},
  \bibinfo{author}{\bibfnamefont{M.}~\bibnamefont{Murphy}},
  \bibinfo{author}{\bibfnamefont{T.}~\bibnamefont{Calarco}},
  \bibinfo{author}{\bibfnamefont{R.}~\bibnamefont{Fazio}},
  \bibinfo{author}{\bibfnamefont{S.}~\bibnamefont{Montangero}},
  \bibinfo{author}{\bibfnamefont{V.}~\bibnamefont{Giovannetti}},
  \bibnamefont{and} \bibinfo{author}{\bibfnamefont{G.~E.}
  \bibnamefont{Santoro}}, \bibinfo{journal}{Physical review letters}
  \textbf{\bibinfo{volume}{103}}, \bibinfo{pages}{240501}
  (\bibinfo{year}{2009}).

\bibitem[{\citenamefont{Deutsch and Jessen}(2010)}]{deutsch2010quantum}
\bibinfo{author}{\bibfnamefont{I.~H.} \bibnamefont{Deutsch}} \bibnamefont{and}
  \bibinfo{author}{\bibfnamefont{P.~S.} \bibnamefont{Jessen}},
  \bibinfo{journal}{Optics Communications} \textbf{\bibinfo{volume}{283}},
  \bibinfo{pages}{681} (\bibinfo{year}{2010}).

\bibitem[{\citenamefont{Giorda et~al.}(2003)\citenamefont{Giorda, Zanardi, and
  Lloyd}}]{Giorda2003}
\bibinfo{author}{\bibfnamefont{P.}~\bibnamefont{Giorda}},
  \bibinfo{author}{\bibfnamefont{P.}~\bibnamefont{Zanardi}}, \bibnamefont{and}
  \bibinfo{author}{\bibfnamefont{S.}~\bibnamefont{Lloyd}},
  \bibinfo{journal}{Phys. Rev. A} \textbf{\bibinfo{volume}{68}},
  \bibinfo{pages}{062320} (\bibinfo{year}{2003}),
  \urlprefix\url{https://link.aps.org/doi/10.1103/PhysRevA.68.062320}.

\bibitem[{\citenamefont{Olschewski}(1972)}]{olschewski1972messung}
\bibinfo{author}{\bibfnamefont{L.}~\bibnamefont{Olschewski}},
  \bibinfo{journal}{Zeitschrift f{\"u}r Physik} \textbf{\bibinfo{volume}{249}},
  \bibinfo{pages}{205} (\bibinfo{year}{1972}).

\bibitem[{\citenamefont{Merkel et~al.}(2008)\citenamefont{Merkel, Jessen, and
  Deutsch}}]{Merkel2008}
\bibinfo{author}{\bibfnamefont{S.~T.} \bibnamefont{Merkel}},
  \bibinfo{author}{\bibfnamefont{P.~S.} \bibnamefont{Jessen}},
  \bibnamefont{and} \bibinfo{author}{\bibfnamefont{I.~H.}
  \bibnamefont{Deutsch}}, \bibinfo{journal}{Phys. Rev. A}
  \textbf{\bibinfo{volume}{78}}, \bibinfo{pages}{023404}
  (\bibinfo{year}{2008}),
  \urlprefix\url{https://link.aps.org/doi/10.1103/PhysRevA.78.023404}.

\bibitem[{\citenamefont{Rabitz et~al.}(2004)\citenamefont{Rabitz, Hsieh, and
  Rosenthal}}]{rabitz2004quantum}
\bibinfo{author}{\bibfnamefont{H.~A.} \bibnamefont{Rabitz}},
  \bibinfo{author}{\bibfnamefont{M.~M.} \bibnamefont{Hsieh}}, \bibnamefont{and}
  \bibinfo{author}{\bibfnamefont{C.~M.} \bibnamefont{Rosenthal}},
  \bibinfo{journal}{Science} \textbf{\bibinfo{volume}{303}},
  \bibinfo{pages}{1998} (\bibinfo{year}{2004}).

\bibitem[{\citenamefont{Hsieh and Rabitz}(2008)}]{Hsieh2008}
\bibinfo{author}{\bibfnamefont{M.}~\bibnamefont{Hsieh}} \bibnamefont{and}
  \bibinfo{author}{\bibfnamefont{H.}~\bibnamefont{Rabitz}},
  \bibinfo{journal}{Phys. Rev. A} \textbf{\bibinfo{volume}{77}},
  \bibinfo{pages}{042306} (\bibinfo{year}{2008}),
  \urlprefix\url{https://link.aps.org/doi/10.1103/PhysRevA.77.042306}.

\bibitem[{\citenamefont{Khaneja et~al.}(2005)\citenamefont{Khaneja, Reiss,
  Kehlet, Schulte-Herbr{\"u}ggen, and Glaser}}]{khaneja2005optimal}
\bibinfo{author}{\bibfnamefont{N.}~\bibnamefont{Khaneja}},
  \bibinfo{author}{\bibfnamefont{T.}~\bibnamefont{Reiss}},
  \bibinfo{author}{\bibfnamefont{C.}~\bibnamefont{Kehlet}},
  \bibinfo{author}{\bibfnamefont{T.}~\bibnamefont{Schulte-Herbr{\"u}ggen}},
  \bibnamefont{and} \bibinfo{author}{\bibfnamefont{S.~J.}
  \bibnamefont{Glaser}}, \bibinfo{journal}{Journal of magnetic resonance}
  \textbf{\bibinfo{volume}{172}}, \bibinfo{pages}{296} (\bibinfo{year}{2005}).

\bibitem[{\citenamefont{Schulte-Herbr{\"u}ggen
  et~al.}(2011)\citenamefont{Schulte-Herbr{\"u}ggen, Sp{\"o}rl, Khaneja, and
  Glaser}}]{schulte2011}
\bibinfo{author}{\bibfnamefont{T.}~\bibnamefont{Schulte-Herbr{\"u}ggen}},
  \bibinfo{author}{\bibfnamefont{A.}~\bibnamefont{Sp{\"o}rl}},
  \bibinfo{author}{\bibfnamefont{N.}~\bibnamefont{Khaneja}}, \bibnamefont{and}
  \bibinfo{author}{\bibfnamefont{S.}~\bibnamefont{Glaser}},
  \bibinfo{journal}{Journal of Physics B: Atomic, Molecular and Optical
  Physics} \textbf{\bibinfo{volume}{44}}, \bibinfo{pages}{154013}
  (\bibinfo{year}{2011}).

\bibitem[{\citenamefont{Viola et~al.}(1999)\citenamefont{Viola, Knill, and
  Lloyd}}]{PhysRevLett.82.2417}
\bibinfo{author}{\bibfnamefont{L.}~\bibnamefont{Viola}},
  \bibinfo{author}{\bibfnamefont{E.}~\bibnamefont{Knill}}, \bibnamefont{and}
  \bibinfo{author}{\bibfnamefont{S.}~\bibnamefont{Lloyd}},
  \bibinfo{journal}{Phys. Rev. Lett.} \textbf{\bibinfo{volume}{82}},
  \bibinfo{pages}{2417} (\bibinfo{year}{1999}),
  \urlprefix\url{https://link.aps.org/doi/10.1103/PhysRevLett.82.2417}.

\bibitem[{\citenamefont{Viola and Lloyd}(1998)}]{PhysRevA.58.2733}
\bibinfo{author}{\bibfnamefont{L.}~\bibnamefont{Viola}} \bibnamefont{and}
  \bibinfo{author}{\bibfnamefont{S.}~\bibnamefont{Lloyd}},
  \bibinfo{journal}{Phys. Rev. A} \textbf{\bibinfo{volume}{58}},
  \bibinfo{pages}{2733} (\bibinfo{year}{1998}),
  \urlprefix\url{https://link.aps.org/doi/10.1103/PhysRevA.58.2733}.

\bibitem[{\citenamefont{Norris et~al.}(2012)\citenamefont{Norris, Trail,
  Jessen, and Deutsch}}]{Noris2012}
\bibinfo{author}{\bibfnamefont{L.~M.} \bibnamefont{Norris}},
  \bibinfo{author}{\bibfnamefont{C.~M.} \bibnamefont{Trail}},
  \bibinfo{author}{\bibfnamefont{P.~S.} \bibnamefont{Jessen}},
  \bibnamefont{and} \bibinfo{author}{\bibfnamefont{I.~H.}
  \bibnamefont{Deutsch}}, \bibinfo{journal}{Phys. Rev. Lett.}
  \textbf{\bibinfo{volume}{109}}, \bibinfo{pages}{173603}
  (\bibinfo{year}{2012}),
  \urlprefix\url{https://link.aps.org/doi/10.1103/PhysRevLett.109.173603}.

\bibitem[{\citenamefont{Blok et~al.}(2021)\citenamefont{Blok, Ramasesh,
  Schuster, O'Brien, Kreikebaum, Dahlen, Morvan, Yoshida, Yao, and
  Siddiqi}}]{Blok2021}
\bibinfo{author}{\bibfnamefont{M.~S.} \bibnamefont{Blok}},
  \bibinfo{author}{\bibfnamefont{V.~V.} \bibnamefont{Ramasesh}},
  \bibinfo{author}{\bibfnamefont{T.}~\bibnamefont{Schuster}},
  \bibinfo{author}{\bibfnamefont{K.}~\bibnamefont{O'Brien}},
  \bibinfo{author}{\bibfnamefont{J.~M.} \bibnamefont{Kreikebaum}},
  \bibinfo{author}{\bibfnamefont{D.}~\bibnamefont{Dahlen}},
  \bibinfo{author}{\bibfnamefont{A.}~\bibnamefont{Morvan}},
  \bibinfo{author}{\bibfnamefont{B.}~\bibnamefont{Yoshida}},
  \bibinfo{author}{\bibfnamefont{N.~Y.} \bibnamefont{Yao}}, \bibnamefont{and}
  \bibinfo{author}{\bibfnamefont{I.}~\bibnamefont{Siddiqi}},
  \bibinfo{journal}{Phys. Rev. X} \textbf{\bibinfo{volume}{11}},
  \bibinfo{pages}{021010} (\bibinfo{year}{2021}),
  \urlprefix\url{https://link.aps.org/doi/10.1103/PhysRevX.11.021010}.

\bibitem[{\citenamefont{Boyd et~al.}(2007)\citenamefont{Boyd, Zelevinsky,
  Ludlow, Blatt, Zanon-Willette, Foreman, and Ye}}]{boyd2007nuclear}
\bibinfo{author}{\bibfnamefont{M.~M.} \bibnamefont{Boyd}},
  \bibinfo{author}{\bibfnamefont{T.}~\bibnamefont{Zelevinsky}},
  \bibinfo{author}{\bibfnamefont{A.~D.} \bibnamefont{Ludlow}},
  \bibinfo{author}{\bibfnamefont{S.}~\bibnamefont{Blatt}},
  \bibinfo{author}{\bibfnamefont{T.}~\bibnamefont{Zanon-Willette}},
  \bibinfo{author}{\bibfnamefont{S.~M.} \bibnamefont{Foreman}},
  \bibnamefont{and} \bibinfo{author}{\bibfnamefont{J.}~\bibnamefont{Ye}},
  \bibinfo{journal}{Physical Review A} \textbf{\bibinfo{volume}{76}},
  \bibinfo{pages}{022510} (\bibinfo{year}{2007}).

\bibitem[{\citenamefont{Gross}(2021)}]{Gross2021}
\bibinfo{author}{\bibfnamefont{J.~A.} \bibnamefont{Gross}},
  \bibinfo{journal}{Phys. Rev. Lett.} \textbf{\bibinfo{volume}{127}},
  \bibinfo{pages}{010504} (\bibinfo{year}{2021}),
  \urlprefix\url{https://link.aps.org/doi/10.1103/PhysRevLett.127.010504}.

\bibitem[{\citenamefont{Brockett}(1973)}]{brockett1973lie}
\bibinfo{author}{\bibfnamefont{R.}~\bibnamefont{Brockett}},
  \bibinfo{journal}{SIAM Journal on Applied Mathematics}
  \textbf{\bibinfo{volume}{25}}, \bibinfo{pages}{213} (\bibinfo{year}{1973}).

\bibitem[{\citenamefont{Schirmer et~al.}(2002)\citenamefont{Schirmer, Solomon,
  and Leahy}}]{schirmer2002degrees}
\bibinfo{author}{\bibfnamefont{S.}~\bibnamefont{Schirmer}},
  \bibinfo{author}{\bibfnamefont{A.}~\bibnamefont{Solomon}}, \bibnamefont{and}
  \bibinfo{author}{\bibfnamefont{J.}~\bibnamefont{Leahy}},
  \bibinfo{journal}{Journal of Physics A: Mathematical and General}
  \textbf{\bibinfo{volume}{35}}, \bibinfo{pages}{4125} (\bibinfo{year}{2002}).

\bibitem[{\citenamefont{Frey et~al.}(2020)\citenamefont{Frey, Norris, Viola,
  and Biercuk}}]{Frey2020}
\bibinfo{author}{\bibfnamefont{V.}~\bibnamefont{Frey}},
  \bibinfo{author}{\bibfnamefont{L.~M.} \bibnamefont{Norris}},
  \bibinfo{author}{\bibfnamefont{L.}~\bibnamefont{Viola}}, \bibnamefont{and}
  \bibinfo{author}{\bibfnamefont{M.~J.} \bibnamefont{Biercuk}},
  \bibinfo{journal}{Phys. Rev. Applied} \textbf{\bibinfo{volume}{14}},
  \bibinfo{pages}{024021} (\bibinfo{year}{2020}),
  \urlprefix\url{https://link.aps.org/doi/10.1103/PhysRevApplied.14.024021}.

\end{thebibliography}
\pagebreak
\onecolumngrid
\appendix
\subsection{Supplemental Material:  Quantum Optimal Control of Nuclear Spin Qudecimals in \textsuperscript{87}Sr}

In this supplement we detail the methods we employ for quantum optimal control.  We consider open loop-control to create arbitrary unitary evolution, a problem which is studied extensively in the literature \cite{jurdjevic1972control,brockett1973lie,schirmer2002degrees, goerz2015optimizing, glaser2015training}. In general consider a Hamiltonian in a $d$-dimension Hilbert space of the form, 
\begin{equation}
H(t)=H_0+\sum_{\lambda=1}^{K}c_\lambda(t)H_{\lambda}.
\label{eqn: control Hamiltonian}
\end{equation}
 The system is controllable if we can generate any $U_0\in SU(d)$ using a set of controls $c_\lambda(t)$,  This means that in a finite time $T$, the Hamiltonian evolution given by the Schr{\"o}dinger equation $\dot{U}=-iH(t)U$, maps the identity operator to any arbitrary unitary operator $U_0$ in the group with arbitrary precision. A necessary and sufficient condition for the controllability is the set of Hamiltonians $\{H_0,H_1,H_2,..., H_K\}$ generate the Lie algebra $\mathrm{su}\left(d\right)$.  
 
 In this letter we consider the control of a nuclear spin $\mathbf{I}$ with dimension $d=2I+1=10$ using a combination of radio-frequency driven Larmor precession and a tensor AC-Stark shift according to the Hamiltonian,
\begin{equation}
H(t)=\Omega_{\mathrm{rf}} \left( \cos[c(t) \pi]I_x+\sin[c(t) \pi] I_y\right)+\beta I_z^2.
\label{eq:Control_Hamiltonian}
\end{equation}
It was proved in \cite{Merkel2008} that by manipulating the phase $c(t)$ the above system is controllable.

We consider two classes of quantum control tasks: preparation of a target pure state $\ket{\psi_\text{tar}}$ and implementation of a target unitary map $U_\text{tar}$ on an arbitrary input state.  We implement these tasks using quantum optimal control. The goal is to find the waveform $c(t)$ which optimizes the objective function.  As a first step we discretize the control waveform as a piecewise constant function over $n$ equal intervals in the time $T$, $\mathbf{c}=\{c_i = c(t_i) | i=1,\dots N\}$.  Optimal control for state preparation and unitary maps follows by maximizing the relevant fidelity, 
\begin{eqnarray}
\mathcal{F}_{\psi}[\bm{c},T]&=&\left|\bra{\psi_{\text{tar}}}U[\bm{c},T]\ket{\psi_0}\right|^2,\\
\mathcal{F}_U[\bm{c},T]&=&\left|\Tr\left(U^{\dagger}_{\text{tar}}U[\bm{c},T]\right)\right|^2/d^2.
\label{eq:fidelity}
\end{eqnarray} 
Here $U[\bm{c},T] =\prod_{i=1}^n e^{-iH(c_i)T/n}$.  To find $\mathbf{c}$, we use the well-known gradient based optimization method GRAPE \cite{khaneja2005optimal}. Robust optimization follows when $T$ is sufficiently large compared to the minimal value $T_*$ set by the quantum speed limit~\cite{caneva2009optimal} and $n$ is sufficiently large compared with the minimal number of parameters necessary to specific the control task.  For a $d$-dimensional Hilbert space, $n_{\min} = 2d-2$ for state preparation and $n_{\min} =d^2-1$ for unitary maps.

 
While in principle we can find simple control waveforms with $n$ close to $n_{\min}$, in practice, the resulting discontinuous waveforms may not be exactly realizable in an experimental implementation. To find waveforms that are more experimentally feasible we constrain the maximum jump allowed between $c_i \text{ and } c_{i+1}$ to create a smoother waveform, as was shown in~\cite{Frey2020}.  Another important ingredient is the choice of the initial seed $\mathbf{c}$ to the GRAPE algorithm. A waveform that yields high-fidelity is not unique, and by choosing smoother initial seed, the optimal solution will be smoother as well.    Here we choose the initial condition where $c_i=0 \hspace{0.1cm}\forall i$.  This is sufficiently small so that the time for computer optimization is reasonable, by sufficiently large that we obtain experimentally feasible waveforms, with a maximum of  $c_{i+1}-c_i \leq 0.4 $.

\begin{figure}
\includegraphics[width=1\textwidth]{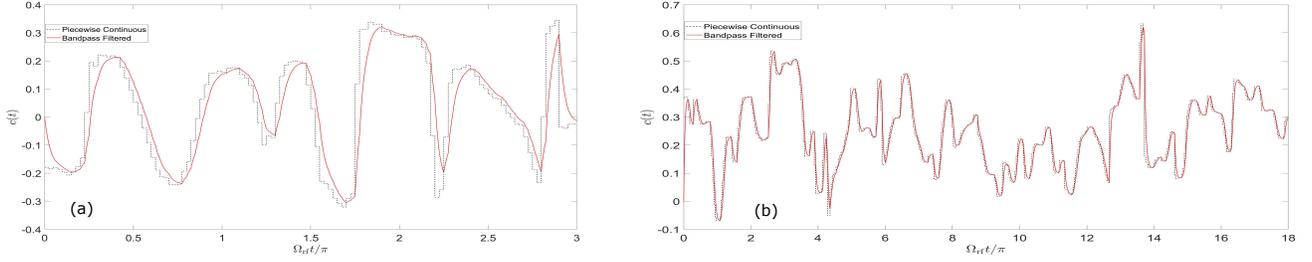}
\caption{Control waveforms for a piecewise constant parameterization, with a limited slew rate (dotted black line) and the waveforms created after the low-pass filter (solid red line ) for the state preparation (a) and unitary mapping (b) with $\Omega_c=10\Omega_{\mathrm{rf}}$.  }
\label{fig:control_waveform_convolution}
\end{figure}

While the quantum control technique described above creates relatively smooth waveforms, there still exist discontinuities which can result in a large slew rate and bandwidth that is outside the range of the physical control. To see how this constraint affects the fidelity, we take simple model to pass the phase waveform through a low-pass filter,
\begin{equation}
     c(t)=\phi(t)/\pi=\Omega_c\int_{0}^{t} c_{ideal}(\xi)\exp\left[-\Omega_c(t-\xi)\right]d\xi\\
    \label{eq:LIS}
\end{equation}
where $c_{ideal}(\xi)$ is the ideal waveform value one would attain as the output of the GRAPE algorithm in a perfect piecewise approach. The waveforms depend on the choice of the corner frequency, $\Omega_c$, which is related to the bandwidth of the controller.  Examples of filtered waveforms obtained using $\Omega_c = 20 \Omega_{\rm{rf}} $ are given in Fig.~(\ref{fig:control_waveform_convolution}).  The resulting waveforms are continuous functions of time and band-limited. 

\begin{figure}[H]
\centering
\includegraphics[width=0.8\textwidth]{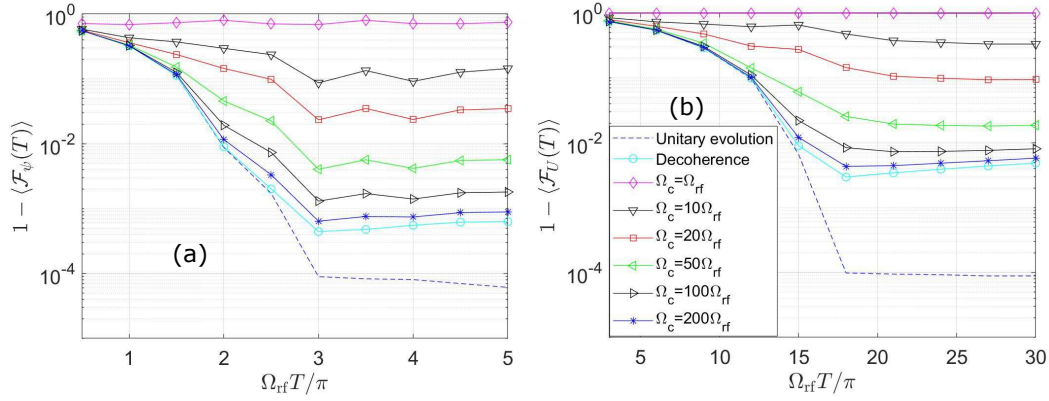}
\caption{The fidelity observed for state preparation (a) and unitary mapping (b) for $\beta=0.4\Omega_{\mathrm{rf}}$ under the full decoherence analysis for different value of the corner frequencey $\Omega_c$.  }
\label{fig:control_waveform_convolution_fidleity}
\end{figure}

The analysis of the control seed after the low-pass filter shows that there is high fidelity operation can be obtain for $\Omega_c \sim 100 \Omega_{\rm{rf}}$, e.g., $\Omega_{\rm{rf}} = 100$ Hz, $\Omega_{c} = 1$ kHz. The decoherence analysis for the continuous waveforms for the state preparation and  unitary mapping (Eq.(6) and Eq.(7) in the main text)  for $\beta=0.4\Omega_{\mathrm{rf}}$ is given in Fig.~(\ref{fig:control_waveform_convolution_fidleity})

\end{document}